\documentclass{article}
	
 \normalbaselines
	\newtheorem{Theorem}{Theorem}[section]

	\newtheorem{Lemma}[Theorem]{Lemma}
	\newtheorem{Corollary}[Theorem]{Corollary}
	
	\newtheorem{Remark}[Theorem]{Remark}

	\newcommand{\Proof}{{\em Proof. }}

	\usepackage{amsmath}
	\usepackage{amsfonts}
	\usepackage[most]{tcolorbox}
	\usepackage{graphicx}

\begin{document}
		
		\title{Asymptotic Normality for the Fourier spot volatility  estimator\\ in the presence of microstructure noise }
		\author{
			Maria Elvira Mancino\footnote{University of Florence, Dept. of Economics and Management, Via delle Pandette 9, 50127 Firenze, Italy:  mariaelvira.mancino@unifi.it}, \  \ Tommaso Mariotti\footnote{Scuola Normale Superiore, Piazza dei Cavalieri 7, 56126 Pisa, Italy: tommaso.mariotti@sns.it}, \  \ Giacomo Toscano\footnote{University of Florence, Dept. of Economics and Management, Via delle Pandette 9, 50127 Firenze, Italy:  giacomo.toscano@unifi.it}
		}

		\maketitle
		
		\textwidth=160mm \textheight=225mm \parindent=8mm \frenchspacing
		\vspace{3mm}

		\begin{abstract}
			
			The main contribution of the paper is proving that the Fourier spot volatility  estimator introduced in \cite{MM} is consistent and asymptotically efficient if the price process is contaminated by microstructure noise. Specifically, in the presence of additive microstructure noise we prove a Central Limit Theorem with the optimal rate of convergence  $n^{1/8}$.
			The result is obtained without the need for any manipulation of the original data or bias correction. Moreover, we complete the asymptotic theory for the Fourier spot volatility  estimator in the absence of noise, originally presented in \cite{Plos}, by deriving a Central Limit Theorem with the optimal convergence rate $n^{1/4}$. Finally, we propose  a novel feasible adaptive method for the optimal selection of the parameters involved in the implementation of the Fourier spot volatility estimator with noisy high-frequency data and  provide support to its accuracy  both numerically and empirically.
		\end{abstract}

		\medskip
		
		
		%MSC: 62G05, 62G20, 60F05, 62P20, 42A38.
		
		\medskip
		
		
		\section{Introduction}
		\label{introduct}
		
		Even though the estimation of the time-varying volatility of financial assets has long been recognized as a relevant topic in financial econometrics (see, e.g.,  \cite{ABDL} and references therein), in the last couple of decades the  increasing availability of high-frequency market data has given an enormous impulse to the investigation of novel estimation methodologies and the exploration of their applications. We refer to \cite{ASJ}  for an extensive review of the relevant literature.
		However, unlike the case of the integrated volatility, the non-parametric  estimation of the instantaneous (or spot) volatility represents a relatively recent topic.
		In this paper we study the asymptotic normality of the non-parametric high-frequency estimator of the instantaneous volatility proposed by \cite{MM}.

		The benchmark for computing the volatility of   an asset  on a fixed time interval with high-frequency data  is provided by the sum of the
		squared intraday returns (i.e., the realized volatility). In the limit as the time
		between two consecutive observations converges to zero, the realized volatility converges to the
		quadratic variation of the asset price process,   and its derivative provides the instantaneous volatility thereof. However, the approximation of the derivative of the quadratic variation  required to obtain the
		spot volatility may generate  appreciable numerical instabilities. Moreover,
		high-frequency data   are affected by the presence of microstructure effects, which cause a discrepancy between asset pricing theory based on semi-martingales and the actual data sampled at very fine intervals.
		
		Kernel smoothing procedures have been extensively adopted to address numerical instabilities and obtain efficient estimators of the spot volatility. Moreover,
		modifications of the   estimators
		have been proposed to correct for the bias  introduced by the presence of microstructure noise. Such modifications, which rely on the pre-processing of price observations, include
		the Two-Scale  Sub-Sampling method by \cite{ZuBo}  and the Pre-Averaging method for kernel-based estimators by \cite{ASJ} and \cite{FigueroaWu}.
		
		This paper deals with an alternative estimation approach, introduced in \cite{MM}, which is based on the Fourier series decomposition of the volatility process. The Fourier method reconstructs the instantaneous volatility as a series expansion, with coefficients gathered from the Fourier coefficients of the price variation.
		Even though the Fourier estimator was originally designed as a global estimator (see also \cite{MM09}), it  also works well  as a point-wise estimator inside the estimation interval.  The main contribution of this paper is to provide a complete study of the point-wise asymptotic normality of the Fourier spot volatility estimator, proving that the latter reaches the optimal rate of convergence and  variance in the absence and in the presence of additive microstructure noise. This result is achieved without any manipulation of the original data or any bias correction, thus supporting the results obtained in a finite-sample setting by \cite{Plos}.

		Several papers have studied the efficiency of the Fourier method in reconstructing   integrated
		volatility and co-volatilities values, even in the presence of microstructure noise, with a   focus on the finite-sample performance, see, e.g., \cite{BarucciReno,HansenLunde,Nielsen,MS2008,MS2011}. The Central Limit Theorem for the Fourier estimator of the integrated covariance was obtained in \cite{MM09}, with a sub-optimal rate of convergence, while  \cite{ClGl}  proved the asymptotic normality with the optimal rate and variance  under the assumption of irregular and non-synchronous observations.
		
		The point-wise Central Limit Theorem for the Fourier spot volatility  estimator in the absence of noise is proved in \cite{Plos}, with a slightly sub-optimal rate of convergence. Moreover, the authors provide an extensive simulation study showing the accuracy of the resulting spot volatility estimates  and their robustness   in the presence of different microstructure noise specifications. A modification of the classical Fourier method is proposed in \cite{CuTe} to obtain a jump-robust estimator of the instantaneous covariance, which is not consistent in the presence of microstructure noise.
		The contribution of various order autocovariances has early been considered by \cite{ZHO} and then by \cite{BNHLS08}. In this spirit, the Fourier spot volatility estimator can be related to a local version of the infinite-lag realized kernels, see \cite{BNHLS08}.
		A different but related study is present in \cite{PaHoLi}. In this study, the authors prove the consistency and asymptotic normality for a class of Fourier-type estimators of the Fourier coefficients  of the covariance (thus integrated quantities), named Fourier Realized Kernels. The Central Limit theorem for any Fourier coefficient holds under some general conditions that allow for microstructure noise effects and asynchronicity between different assets. The rate of convergence depends on the relative liquidity between assets and reaches the value $n^{1/5}$ if the assets are synchronous. However, the rate of convergence of the spot Fourier Realized Kernel estimator is not investigated.
		Finally, a Fourier-type spot volatility estimator is considered also in \cite{MMR}. Nonetheless, the estimator studied in \cite{MMR}, Proposition 4.2, substantially differs from the Fourier spot volatility estimator, precisely being a F\'{e}j\`{e}r kernel-based realized estimator. In fact, the authors get rid of the cross products, which are the main by-product of the original convolution formula (see \cite{MM09}), thus loosing the robustness of the Fourier estimator with respect to microstructure noise.
		
		In this paper, we first  fine-tune the result of \cite{Plos}, that is, under the assumption that the variance process is also a continuous It\^o semimartingale, we prove that the Fourier spot volatility estimator attains the optimal rate of convergence, $n^{1/4}$. Secondly, we prove that, in the presence of additive microstructure noise, the rate of convergence attained by the Fourier estimator is optimal, that is, $n^{1/8}$. This is the convergence rate of the spot volatility estimators in \cite[Section 8.7]{ASJ} and \cite[Theorem 2.2]{FigueroaWu}, while the Two-Scale realized spot variance in \cite[Theorem 2]{ZuBo} attains the convergence rate $n^{1/12}$. However, while  all these estimators require a preliminary manipulation of price observations and a bias-correction, the Fourier estimator does not need any modification to be robust to microstructure noise.
		In fact, the efficient implementation of the Fourier spot volatility estimator requires the balancing of three frequencies: the number of log-return observations $n$; the cutting frequency in the convolution formula that yields the Fourier coefficients of the volatility; the number of Fourier coefficients of the volatility to be used in the Fourier inversion formula (respectively, $N$ and $M$ throughout the paper).
		In the presence of noise, the effect of the latter is ignored by cutting out the highest frequencies in the construction of the Fourier coefficients of the volatility, thereby leading to a noise-robust estimator of the instantaneous volatility. Specifically, while in the absence of noise (see Theorem \ref{TLC1}) the rate-variance optimal choice of the cutting frequency $N$ is the Nyquist frequency $n/2$, in the presence of noise (see Theorem \ref{TLCnoise}) it is optimal to select a smaller $N$, compared to the no-noise case, namely $N=O(n^{1/2})$.  The frequency $M$ is optimized as $M=O(N^{1/2})$ in both the absence and the presence of noise.

		In this paper, we also conduct a  simulation study on the finite-sample performance of the Fourier instantaneous volatility estimator with high-frequency noisy data
		and provide a refinement of the numerical results presented in \cite{Plos} by using as guidance the newly derived asymptotic theory. Specifically, the aim of the simulation study is three-fold.

		First, we investigate  the relationship between the efficient selection of the parameters $N$ and $M$ and the intensity of the noise.  The rate-optimal Central Limit Theorem  (see Theorem \ref{TLCnoise}) assumes that $N\sim cn^{1/2}, \, c>0,$ and  $M \sim aN^{1/2}, \, a>0, $  as $n,N,M \to \infty$. Thus, the selection  of $N$ and $M$ may reduce to  choosing the values of the constants $c$ and $a$.  For what concerns $c$,  simulations suggest that it should be carefully decreased   in correspondence of the increase of the noise intensity   to preserve the accuracy of the estimator. As for  $a$,   numerical evidence instead suggests that its optimal value  is  rather insensitive to the noise level. These findings provide a more robust insight into the sensitivity of the efficiency of the Fourier estimator compared to the numerical study in \cite{Plos}, where the authors  assumed $N \sim (1/2) n^\alpha$ and $M \sim (1/16\pi)  n^\beta$  and attempted to optimize the rates $(\alpha, \beta)$ numerically, in correspondence of different noise intensities and specifications.

		Secondly, we introduce a novel feasible adaptive method to select the parameters  $N$ and $M$ from market data and  evaluate its efficiency numerically. The method   is based on the gradient-descent algorithm and optimizes the mean integrated squared error. Under the assumptions that ensure  asymptotic normality of the estimation error, the mean integrated squared error  can be approximated  by the integrated asymptotic error variance at high-frequencies (see \cite[Section 3.4]{ZuBo}). So far, to the best of our knowledge, the only feasible method available in the literature for selecting $N$ and $M$ was the indirect inference procedure described in \cite[Section 4.2]{Plos}, based on the testing of the empirical distribution of the log-returns standardized by volatility estimates.  The integrated asymptotic error variance of the Fourier estimator  in the rate-efficient case depends on the integrals of the variance, quarticity and volatility of volatility processes, along with the  variance of the noise (see Theorem \ref{TLCnoise}). The aforementioned integrated quantities   can be estimated  by means of non-parametric Fourier estimators (see, resp.,   \cite{MS2008}, \cite{MS2012} and \cite{scm}), while the noise variance   can be estimated using the estimator proposed by \cite{BR05}.  Simulations show that  the   adaptive method introduced in this paper yields a very satisfactory performance. Moreover, its accuracy   is supported also by an empirical exercise, conducted with high-frequency prices of the S\&P500  index sampled over the period March, 2018 - April, 2018.

		Finally,  we compare the performance of  the Fourier estimator with that of alternative noise-robust spot volatility estimators. In particular,  the comparison conducted in this paper extends the one presented in \cite{Plos} by taking into account the recently proposed Pre-averaging kernel estimator by \cite{FigueroaWu}. Numerical results suggest  that the Fourier estimator  may have a competitive edge w.r.t. the alternative noise-robust  estimators,  in line with   \cite{Plos}.  However, it is worth underlining that the performance gap w.r.t. the Pre-averaging kernel estimator  is smaller in scenarios characterized by higher noise. The findings of the comparative study conducted in this paper are coherent with those of the simulation study by \cite{MLT}, where  the data-generating process is the limit order book simulator by \cite{HLR}, which is able to reproduce several microstructural characteristics of financial markets.
		
		The paper is organized as follows. Section \ref{SVFE}  contains definitions and assumptions. Asymptotic results in  the absence and the presence of microstructure noise are detailed, resp., in Sections \ref{NONOISE}   and \ref{AsymptoticNoise}. Sections \ref{sim}   illustrates   the simulation study  and Section \ref{emp} contains  the empirical exercise. Section \ref{concl} concludes.  Appendix A contains the proofs, while   Appendix B resumes some auxiliary  results on the  F\'{e}j\`{e}r and Dirichlet kernels.

		\section{Fourier estimator of spot volatility}
		\label{SVFE}
		
		In this section we introduce the setting and   recall the definition of the Fourier estimator of the spot volatility{\footnote{Hereinafter, we will follow the relevant econometric literature by using the term volatility as a synonym of variance,
				thus referring to $\sigma^2(t)$ as the volatility process. We will do the same for the volatility of volatility.}} originally proposed in \cite{MM}.
		The Fourier estimation method consists in a two-step procedure: first the Fourier coefficients of the volatility process are estimated and secondly the volatility path is reconstructed using the Fourier-F\'{e}j\`{e}r inversion formula.
		
		The following assumptions are made.
		
		\medskip
		{\rm (A.I)} The price process $p$ is a continuous It\^o semimartingale satisfying the stochastic differential equation
		$$dp(t)= \sigma(t)\,dW_t + b(t)\,dt,$$
		where $W$ is a standard Brownian motion on a filtered probability space
		$(\Omega, ({\cal F}_t)_{t\in [0,T]}, P)$ satisfying the usual conditions.
		
		\par
		{\rm (A.II)} The spot volatility process $\sigma^2$ is an It\^o semimartingale
		$$d\sigma^2(t)= \gamma(t) dZ_t + b_{v}(t) dt,$$
		where $Z$ is a standard Brownian motion adapted to the filtration
		${\cal F}$ and such that $d\langle W,Z\rangle_t= \rho\, dt $, with constant $\rho$.

		\par
		{\rm (A.III)}
		The processes $\sigma$, $b$, $\gamma$ and $b_{v}$  are
		continuous and adapted stochastic processes defined on the same probability space
		$(\Omega, ({\cal F}_t)_{t\in [0,T]}, P)$ and such that, for any $p\geq 1$,
		$$
		E\left[\int_0^T \sigma^p(t)dt\right]< \infty \ , \ E\left[\int_0^T b^p(t) dt\right]<\infty \ , \ E\left[\int_0^T \gamma^p(t)dt\right]< \infty \ , \ E\left[\int_0^T b_v^p(t)dt\right]< \infty.
		$$
		The processes are specified in such a way that $\sigma$ and $\gamma$ are a.s. positive.
		
		By changing  the origin of time and scaling the unit of time, we can always reduce ourselves to the case where the time window $[0,T]$ becomes $[0,2\pi]$.
		Suppose that the asset log-price $p$ is observed at discrete, irregularly-spaced points in time:
		$\{ 0=t_{0,n} \leq  \ldots t_{i,n} \ldots \leq t_{n,n}=2\pi \}$.
		For simplicity, we will omit the second index $n$. Let $\rho(n):= ~ \max_{0\leq h \leq n-1}|t_{h+1}-t_{h}|$ and suppose that $\rho(n) \to 0$ as $n\to \infty$.
		
		\medskip
		
		Consider the following interpolation formula
		$$
		p_{n}(t):= ~ \sum_{i=0}^{n-1} p(t_i) I_{[t_i,t_{i+1}[}(t)
		$$
		and the discrete Fourier coefficients of $dp$
		$$c_k(dp_n):= \frac{1}{2\pi}\sum_{j=0}^{n-1} e^{-{\rm i}t_j k} \delta_j(p),$$
		with $\delta_i(p):=p(t_{i+1})-p(t_i)$.
		According to \cite{MM09}, for any $t\in (0,2\pi)$, the spot volatility estimator is defined as follows
		\begin{equation}
			\label{StimatoreDiscreto1}
			\widehat \sigma^2_{nNM}(t):= \sum_{|k|\leq M} \left(1- \frac{|k|}{ {M+1}}\right) e^{{\rm i}tk} c_k(\sigma^2_{nN}),
		\end{equation}
		where $c_k(\sigma^2_{nN})$ is an unbiased estimator of the $k$-th Fourier coefficient of the volatility process, obtained through the convolution formula:
		\begin{equation}
			\label{convo}
			c_k(\sigma^2_{nN}):= \frac{2\pi}{2N+1} \sum_{|h|\leq N} c_h(dp_n)c_{k-h}(dp_n).
		\end{equation}
		The estimator (\ref{StimatoreDiscreto1}) is a consistent estimator of $\sigma^2(t)$ for $t\in (0,2\pi)$, as proved in Theorem \ref{TLC1}. On the contrary, for  $t=0$ or $t=2\pi$, which represent fixed times of discontinuity, the estimator converges to $(\sigma^2(t^-)+\sigma^2(t))/2$.
		
		It is possible to express (\ref{convo}) by means of the rescaled Dirichlet kernel $D_N$, defined in (\ref{dirichlet}), as follows:
		$$
		c_k(\sigma^2_{nN})=\frac{1}{2\pi} \sum_{i=0}^{n-1}\sum_{j=0}^{n-1}   D_N(t_j-t_i) e^{-{\rm i}k t_j}\delta_i(p)\delta_j(p).
		$$
		Therefore, the Fourier spot volatility estimator
		(\ref{StimatoreDiscreto1}) can be written with two kernels as follows:
		\begin{equation}
			\label{StimatoreDiscreto}
			\widehat \sigma^2_{nNM}(t)= \frac{1}{2\pi} \sum_{i=0}^{n-1}\sum_{j=0}^{n-1}  F_M(t-t_j)  D_N(t_j-t_i) \delta_i(p)\delta_j(p),
		\end{equation}
		where $F_M$ is the F\'{e}j\`{e}r kernel defined in (\ref{FejerDef}).
		
		\begin{Remark}
			If the auto-covariances in the convolution formula (\ref{convo}) are ignored, the Fourier estimator (\ref{StimatoreDiscreto}) becomes a kernel-type spot volatility estimator (the kernel being the F\'{e}j\`{e}r one), namely it reduces to
			\begin{equation}
				\label{StimatoreDiscreto2}
				\widehat \sigma^2_{n M}(t) =\frac{1}{2\pi} \sum_{j=0}^{n-1}  F_M(t-t_j) (\delta_j(p))^2.
			\end{equation}
			As the inversion formula in the definition (\ref{StimatoreDiscreto1}) can be obtained with different kernels, the Fourier spot volatility estimator generalizes   kernel-type estimators for the joint presence of two kernels.
			Kernel-type spot volatility estimators in the absence of noise are studied in \cite{Krist,MMR,FigueroaLi}. However, kernel-type estimators are not robust in the presence of noise, as shown in \cite{Plos}. Therefore, the authors need to employ the pre-averaging technique and introduce a bias correction in the noise-contaminated case, see \cite{ZuBo,FigueroaWu}.
			
			On the contrary, in Section \ref{AsymptoticNoise}  we will prove that the presence of the cross-products, arising from the convolution formula in the original definition of the Fourier estimator (\ref{StimatoreDiscreto1}), is crucial to ensure the robustness of the estimator
			in the presence of microstructure noise, without the need of any manipulation of data, like sparse sampling or preaveraging, and without resorting to any bias correction.  In the context of integrated volatility estimators, the contribution of various order auto-covariances has early been considered by \cite{ZHO} and \cite{BNHLS08} to correct for the bias of the realized-variance-type estimators in the presence of noise.
			The Fourier estimator remains unbiased and efficient with an appropriate choice of the cutting frequency $N$, which controls the convolution.
		\end{Remark}

		\begin{Remark}
			\label{data}
			The definition of the Fourier spot volatility estimator does not require to consider equidistant observations. However, for simplicity, in the next sections we will refer to the latter sampling scheme. The general case can be obtained as in \cite{MM09}.
		\end{Remark}

		\section{Asymptotic Normality in the absence of microstructure noise}
		\label{NONOISE}
		
		In this section we study the asymptotic normality of the Fourier spot volatility estimator defined in (\ref{StimatoreDiscreto1}). We prove that the estimator reaches the optimal convergence rate, $n^{1/4}$, as well as the efficient variance. This result completes the Central Limit Theorem  with slightly sub-optimal rate proved  for the Fourier spot volatility  estimator (\ref{StimatoreDiscreto1}) in \cite{Plos}  under the hypothesis that the volatility process is a random function with H\"older continuous paths.

		\begin{Theorem}
			\label{TLC1}
			Under the assumptions $(A.I)-(A.II)-(A.III)$ and the condition $\lim_{n,N\to \infty}{N / n}= c>0$,  for any fixed $t \in (0,2\pi)$, as $n,N,M \to \infty$, the following  stable convergence in law holds:
			
			\medskip\noindent
			- if $\lim_{n,M\to \infty}{M n^{-1/\tau}}= a>0$, for $1<\tau<2$,
			\begin{equation}
				\label{NOTEF}
				n^{1/2}M^{-1/2} \left( \widehat \sigma^2_{nNM}(t)-  \sigma^2(t) \right) \to  {\cal N} \left(0, \frac{4}{3} (1+2K(2c)) \, \sigma^4(t)\right),
			\end{equation}
			\medskip\noindent
			- if $\lim_{n,M\to \infty}{M n^{-1/2}}= a>0$,
			\begin{equation}
				\label{EF}
				n^{1/2}M^{-1/2} \left( \widehat \sigma^2_{nNM}(t)-  \sigma^2(t) \right) \to  {\cal N} \left(0,  {4\over 3} (1+2K(2c)) \sigma^4(t)+  {2\pi \over {3a^2}} \, \gamma^2(t) \right),
			\end{equation}
			where the constant $K(c)$ is defined by
			\begin{equation}
				\label{etaci}
				K(c):=\frac{1}{2c^2}r( c)(1-r( c)),
			\end{equation}
			with
			$r(x)=x-[x]$, with $[x]$ denoting the integer part of $x$.
		\end{Theorem}

		\begin{Remark}
			\label{Rate}
			{\sl (Convergence Rate)}
			\par\noindent
			The asymptotic result (\ref{NOTEF}) is in line with the one obtained in \cite{Plos}, where the rate is optimal up to a logarithmic correction, that is, is equal to $n^{1/4} (\log n)^{-1/2}$.
			In \cite{Plos} it is assumed that ${N/ n} \sim c>0$ and ${M n^{-1/\tau}}\sim a>0$, for $1<\tau<2$. Note that in this case $Mn^{-1/2} \to \infty$.
			It is worth noting that the Central Limit Theorem  in \cite{Plos} has been proved for a volatility process $\sigma^2$ belonging to a more general class, that is, under the assumption that
			$\sigma^2$ is a.s. H\"older continuous in $[0,2\pi]$ with parameter $\alpha \in (0,1/2)$.
			\par
			In comparison with Theorem 8.6 \cite{ASJ}, letting $\beta= \sqrt{2\pi}/a$, then the asymptotic result (\ref{NOTEF}) corresponds to the case $\beta=0$, where the rate is sub-optimal. The asymptotic result (\ref{EF}) corresponds to the case $\beta \in (0,\infty)$, which instead attains the optimal rate equal to $n^{1/4}$. An analogous result is obtained in Theorem 2.1 \cite{FigueroaWu}, where alternative localization kernels are considered.
		\end{Remark}
		
		\begin{Remark}
			\label{OptVar}
			{\sl (Asymptotic Variance)}
			\par\noindent
			For what concerns the asymptotic error variance, we observe that the choice $c=1/2$ in (\ref{etaci}) gives $K(2c)=0$. The computation of $K(2c)$ is
			obtained in \cite{ClGl} and takes into account the behaviour of the discretized Dirichlet kernel if $N$ has the same order of the observation grid points $n$. Note that $K(2c)$ is  nonnegative for any positive $c$ and equal to zero when $c=(1/2)k$, $k=1,2,\ldots$.
			The choice $k=1$  (i.e.,  $c=1/2$) corresponds to the natural choice of the Nyquist frequency $n/2$ for $N$
			and provides  the optimal asymptotic variance with regards to the component $(4/3)\sigma^4(t)$, which results smaller than its counterpart $2\sigma^4(t)$ in Theorem 8.6 by \cite{ASJ}. Further, letting $\beta= \sqrt{2\pi}/a$, even the second addend of the asymptotic variance in the efficient-rate case (\ref{EF}), that is, $2\pi/(3a^2)\gamma^2(t)$, is  smaller than the corresponding term $2\pi/a^2 \gamma^2(t)$ in   Theorem 8.6 by \cite{ASJ}.
			This is due to the presence of the F\'{e}j\`{e}r kernel in the Fourier inversion formula, see also \cite{CuTe}.
		\end{Remark}
		
		We stress that with the choice of $c=1/2$ (in other words, of $N$ equal to the Nyquist frequency) the Fourier estimator has the same rate of convergence and asymptotic variance of the F\'{e}j\`{e}r kernel-based realized spot volatility (\ref{StimatoreDiscreto2}) considered in \cite{Krist,MMR}. However, their estimator is not robust to market microstructure and the application of a pre-averaging procedure with a bias-correction is needed.
		In summary, with an appropriate choice of $N/n$, the effect of adding the cross terms in (\ref{StimatoreDiscreto}), which is essential in order to get an noise-robust estimator, is also non-detrimental in view of the asymptotic efficiency.

		Based on the previous remarks, the following result easily follows.
		
		\begin{Corollary}
			Under the assumptions $(A.I)-(A.II)-(A.III)$ and the conditions $\lim_{n,N\to \infty}{N / n}= {1 /2}$ and  $\lim_{n,M\to \infty}{M n^{-1/\tau}}= a>0$, for $1<\tau<2$,
			for any fixed $t \in (0,2\pi)$, the following stable convergence  in law holds:
			$$
			n^{1/2}M^{-1/2}\left( \widehat \sigma^2_{nNM}(t)-  \sigma^2(t) \right) \to  {\cal N} \left(0, {4\over 3}\, \sigma^4(t)\right).
			$$
		\end{Corollary}

		\section{Asymptotic Normality in the presence of microstructure noise}
		\label{AsymptoticNoise}

		In this section we study the asymptotic normality for the estimator (\ref{StimatoreDiscreto1}) when the price process is contaminated by microstructure noise.
		The main result of this section shows that the spot volatility estimator (\ref{StimatoreDiscreto1}) reaches the optimal rate of convergence and asymptotic variance.

		\medskip
		
		The following assumption for the noise process is considered:
		
		\medskip
		
		(N) The noise process $\eta$ is such that $(\eta_{t_i})_{i\geq 0}$ is a family of i.i.d. random
		variables, independent from price process $p$, and for any $t_i$ it holds:
		$$\begin{cases}
			E[\eta_{t_i}]=0, \\
			\xi:=E[\eta_{t_i}^2]<  \infty ,\\
			\omega :=E[\eta_{t_i}^4]<  \infty. \\
		\end{cases}$$
		Let $\varepsilon_i:=\eta_{i+1}-\eta_i$, where for simplicity $\eta_i:=\eta_{t_i}$.
		
		\bigskip
		
		Denote by $\widetilde\sigma_{nNM}^2(t)$ the Fourier estimator (\ref{StimatoreDiscreto1}) obtained by using noise contaminated returns, that is,
		\begin{equation}
			\label{StimatoreDiscreto22}
			\widetilde \sigma^2_{nNM}(t)= \sum_{|k|\leq M} \left(1- {|k|\over {M+1}}\right) e^{{\rm i}tk} c_k(\widetilde\sigma^2_{nN}),
		\end{equation}
		where $c_k(\widetilde\sigma^2_{nN})$ is the estimator of the $k$-th Fourier coefficient of the volatility
		obtained via (\ref{convo}) using the contaminated prices $\widetilde p_{t_i}:= p_{t_i}+ \eta_{t_i}$.

		\begin{Theorem}
			\label{TLCnoise}
			Under the assumptions $(A.I)-(A.II)-(A.III)-(N)$ and the condition $\lim_{N,n\to \infty}{N  {n^{-1/2}}}= c>0$,   as $n,N,M \to \infty$,  for any fixed $t \in (0,2\pi)$, the following stable convergence in law holds:
			
			\medskip\noindent
			- if $\lim_{N,M\to \infty}{M \, N^{-1/\tau}}= a>0$, for $1<\tau<2$,
			\begin{equation}
				\label{NOTEFm}
				n^{1/4}M^{-1/2}\left( \widetilde \sigma^2_{nNM}(t)-  \sigma^2(t) \right) \to  {\cal N} \left(0, {1\over c}{2\over 3}\sigma^4(t)+c {2\pi \over 9} \sigma^2(t) \xi+  c^3 {4\pi^2 \over {15}} \xi^2\right),
			\end{equation}
			\medskip\noindent
			- if $\lim_{N,M\to \infty}{M \, N^{-1/2}}= a>0$,
			\begin{equation}
				\label{EFm}
				n^{1/4}M^{-1/2}\left( \widetilde \sigma^2_{nNM}(t)-  \sigma^2(t) \right) \to  {\cal N} \left(0, {1\over c}{2\over 3} \sigma^4(t) + {1\over {a^2c}} \,{2\pi \over 3}\,  \gamma^2(t) + c {2\pi \over 9} \sigma^2(t) \xi+ c^3 {4\pi^2 \over {15}} \xi^2\right).
			\end{equation}
		\end{Theorem}

		\begin{Remark}
			\label{ratenoise}
			{\sl (Convergence rate)}
			\par\noindent
			In the presence of microstructure noise, the convergence rate attained by the Fourier estimator is optimal in (\ref{EFm}), that is $n^{1/8}$, as long as $Nn^{-1/2} \sim c>0$ and $MN^{-1/2}\sim a>0$. This is the rate found for the spot volatility estimators in \cite[Section 8.7]{ASJ}   and in \cite[Theorem 2.2]{FigueroaWu}. However, while in these cases pre-averaging and a bias-correction are needed, the Fourier estimator (\ref{StimatoreDiscreto1}) does not need any modification.
			The case (\ref{NOTEFm}) reaches a slightly sub-optimal rate, if $M N^{-1/\tau}\sim a>0$, for $1<\tau<2$; however, the asymptotic variance is smaller and does not depend on the volatility of volatility.
		\end{Remark}
		
		\begin{Remark}
			\label{variancenoise}
			{\sl (Asymptotic Variance)}
			\par\noindent
			The asymptotic variance in (\ref{EFm}) depends on the spot quarticity, the spot volatility of volatility, and the variance of the noise. All these factors can be estimated, thus in principle a feasible Central Limit Theorem  is achievable. The same addends appear in the asymptotic variance of the pre-averaging and bias-corrected kernel-type estimators by \cite{ASJ,FigueroaWu}.
			Notice that, as here the condition $N/n\to 0$ is in force, the quantity $K(2c)$ in (\ref{etaci}), which is due to the Dirichlet kernel discretization, disappears (see also Lemma \ref{Key1}).
		\end{Remark}
		
		It is worth noticing the role of the cutting frequency $N$, which is evident by comparing   Theorems \ref{TLCnoise} and \ref{TLC1}. In Theorem \ref{TLCnoise} the frequency $N$ is chosen smaller with respect to the no-noise case, in line with the findings of \cite{MS2008}. In fact, the role of the cutting frequency $N$ is that of filtering the microstructure noise: the high-frequency noise or short-run noise is ignored by cutting the highest frequencies in the construction of the  Fourier coefficients of the volatility $c_k(\sigma^2_{nN})$.

		\section{Simulation study}\label{sim}

		In this section we illustrate   a simulation study
		on the finite-sample performance of the Fourier estimator of the spot volatility (\ref{StimatoreDiscreto22}). The aim  of the simulation study is three-fold.  First, we  investigate how the optimal values of the parameters $N$ and $M$ vary in correspondence of different noise intensities. Secondly, we introduce a  feasible adaptive procedure  for the optimal selection of  $N$ and $M$, and numerically evaluate its efficacy. Finally, we compare the finite-sample efficiency of the Fourier estimator  with that of   alternative noise-robust   spot volatility estimators.
		
		\subsection{Simulation design}
		
		For the simulation study, we generated  discrete 1-second observations from two models  satisfying Assumptions (A.I), (A.II) and (A.III): the One Factor Stochastic Volatility model (commonly referred to as SV1F, see, e.g., \cite{HuaTau}) and the model by \cite{heston}. Note that the former model is adopted in the simulation exercises by \cite{ZuBo}, \cite{Plos} and \cite{FigueroaWu}, while the latter is used in the numerical studies by \cite{Plos}.
		
		The SF1V model reads as

		\begin{equation}
			\begin{cases}
				dp(t) &=    \sigma(t) dW_t +\mu    dt\\
				\sigma(t) &=\textrm{e}^{\beta_0+\beta_1 \tau(t)}\\
				d\tau(t)&=  dZ_t + \alpha \tau(t)
			\end{cases}
		\end{equation}
		where $W$ and $Z$ are two Brownian motions with correlations $\rho$. The parameter vector used for the simulations is  $(\mu,  \beta_1, \alpha, \beta_0,\rho) = (0.03,0.125,-0.025,{\beta_1}/({2\alpha}), -0.3).$  Moreover, we set $p(0)=\textrm{ln}(100)$, while $\tau(0)$ is drawn from the stationary distribution of $\tau$, that is, from $\mathcal{N}(0,-{1}/({2\alpha}))$.
		
		The dynamics of the Heston model instead  follow
		\begin{equation}
			\begin{cases} dp(t) &=    \sigma(t) dW_t+ \left(\mu-\frac{1}{2}\sigma^2(t)\right) dt\\ d\sigma^2(t) &= \gamma \sigma(t)  dZ_t+\theta (\alpha-\sigma^2(t))dt ,
			\end{cases}
		\end{equation}
		where, again, $W$ and $Z$ are two Brownian motions with correlations $\rho$.  For the simulations, we set $(\mu, \theta, \alpha, \gamma, \rho) = (0.001, 0.3, 0.002, 0.03, -0.5).$  Further, we set $p(0)=\textrm{ln}(100)$, while $\sigma^2(0)$ is drawn from the stationary distribution of $\sigma^2$, i.e., from $\Gamma\left(u_1= \frac{2\theta \alpha}{\gamma^2},u_2=\frac{\gamma^2}{2\theta}\right)$, where $u_1$ and $u_2$ denoted, resp., the shape and scale parameters.

		The additive noise process $\eta$ was simulated as a sequence of i.i.d. zero-mean Gaussian random variables with variance equal to  $\xi$  and satisfying
		
		$$\sqrt{\xi} = \zeta \, std(r), $$
		where $std(r)$ is the   standard deviation of the noise-free 1-second log-returns and
		$\zeta$ denotes the noise-to-signal ratio.  Specifically, we used three increasing values of   $\zeta$, namely $\zeta=1,2,3$. For each data-generating process and each level of the noise-to-signal ratio, we simulated 1000 trajectories with horizon $T$ equal to one day, corresponding to 6.5 hours.

		\subsection{Sensitivity of the optimal $N$ and $M$ to different noise intensities}\label{sect:sensitivity}
		
		In this subsection we investigate how the optimal values of the parameters $N$ and $M$  change in correspondence of changes in the intensity of the noise. For the implementation of the Fourier spot volatility estimator\footnote{The MATLAB code used for the implementation of the estimator is available in Appendix B.4   of \cite{book}.}, we used the highest available sampling  frequency, that is, $\rho(n)$ equal to one second, corresponding to the sample size  $n=23400$. Further, based on the conditions of the rate-optimal Central Limit Theorem  in (\ref{EFm}), we set $N = \lfloor c \, \sqrt{ n} \rfloor$ and $M =  \lfloor  a \, \sqrt{N}\rfloor$, and  evaluated the sensitivity of the finite-sample performance to changes in the constants $c$ and $a$. Specifically, we allowed ${c}$ and ${a}$ to vary, resp., in the range $[1,10]$ and $[0.1,0.5]$, with a step size equal to, resp., $1$ and $0.1$. Following \cite[Section 4.2]{Plos}, the spot volatility was reconstructed every minute.
		
		The finite-sample performance over each simulated trajectory was assessed based on the mean integrated  squared error (MISE) and  mean integrated absolute error (MIAE), which are defined as
		$$ MISE:=E\left[\int_0^T \Big(\widetilde \sigma^2_{n N M}(t)- \sigma^2(t) \Big)^2  dt \right] $$
		and
		$$ MIAE:=E\left[\int_0^T \Big|\widetilde \sigma^2_{n N M}(t)- \sigma^2(t) \Big| dt\right].  $$
		Note that we measured time in days, so  all the results appearing hereinafter were obtained for $T=1$.
		\par
\begin{figure}[h!]
			\includegraphics[width=\textwidth]{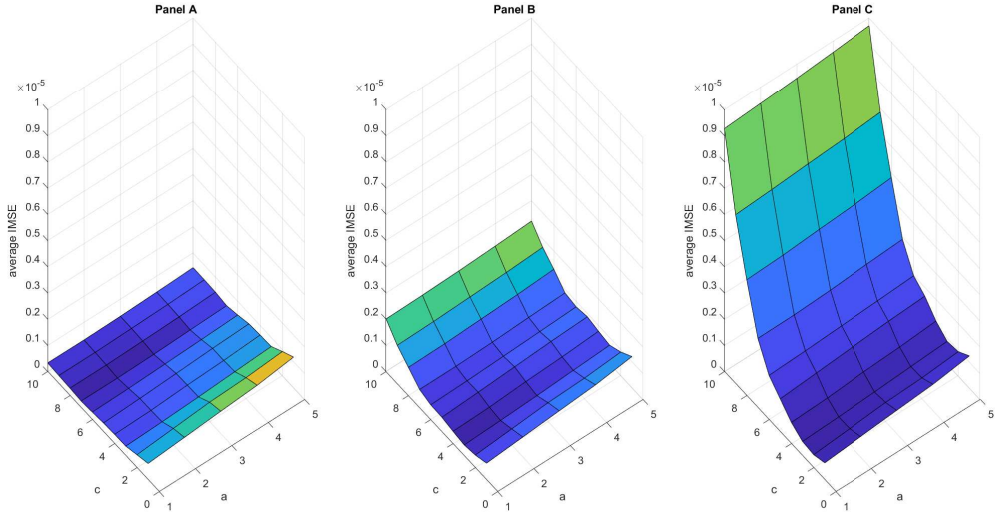}
			\caption{MISE values over 1000 simulated trajectories of the SV1F model with $\zeta=1$ (Panel A), $\zeta=2$ (Panel B) and $\zeta=3$ (Panel C).}\label{fig:IMSEexp}
		\end{figure}
\begin{figure}[h!]
			\includegraphics[width=\textwidth]{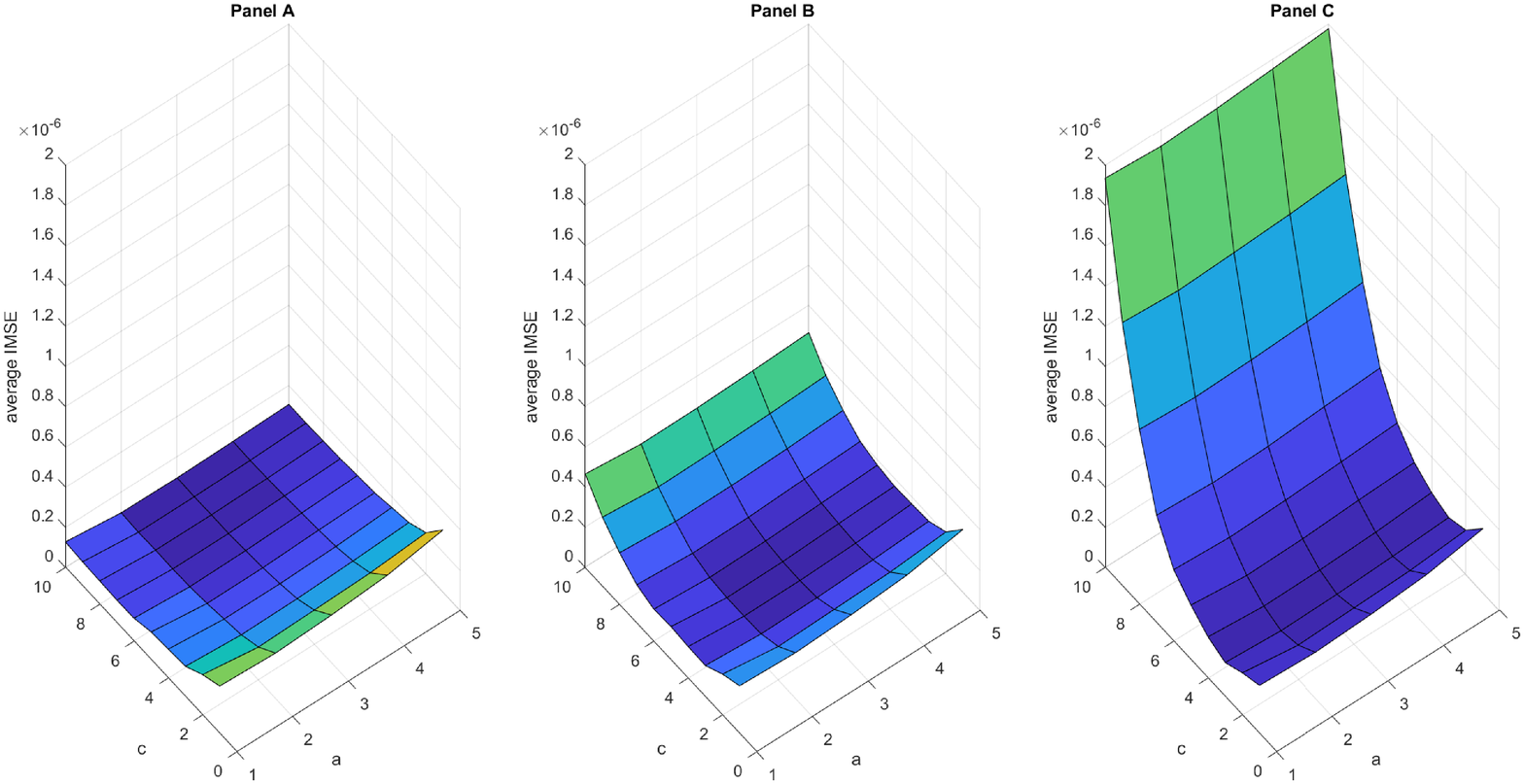}
			\caption{MISE values over 1000 simulated trajectories of the Heston model with $\zeta=1$ (Panel A), $\zeta=2$ (Panel B) and $\zeta=3$ (Panel C). }\label{fig:IMSEhes}
		\end{figure}
\begin{figure}[h!]
			\includegraphics[width=\textwidth]{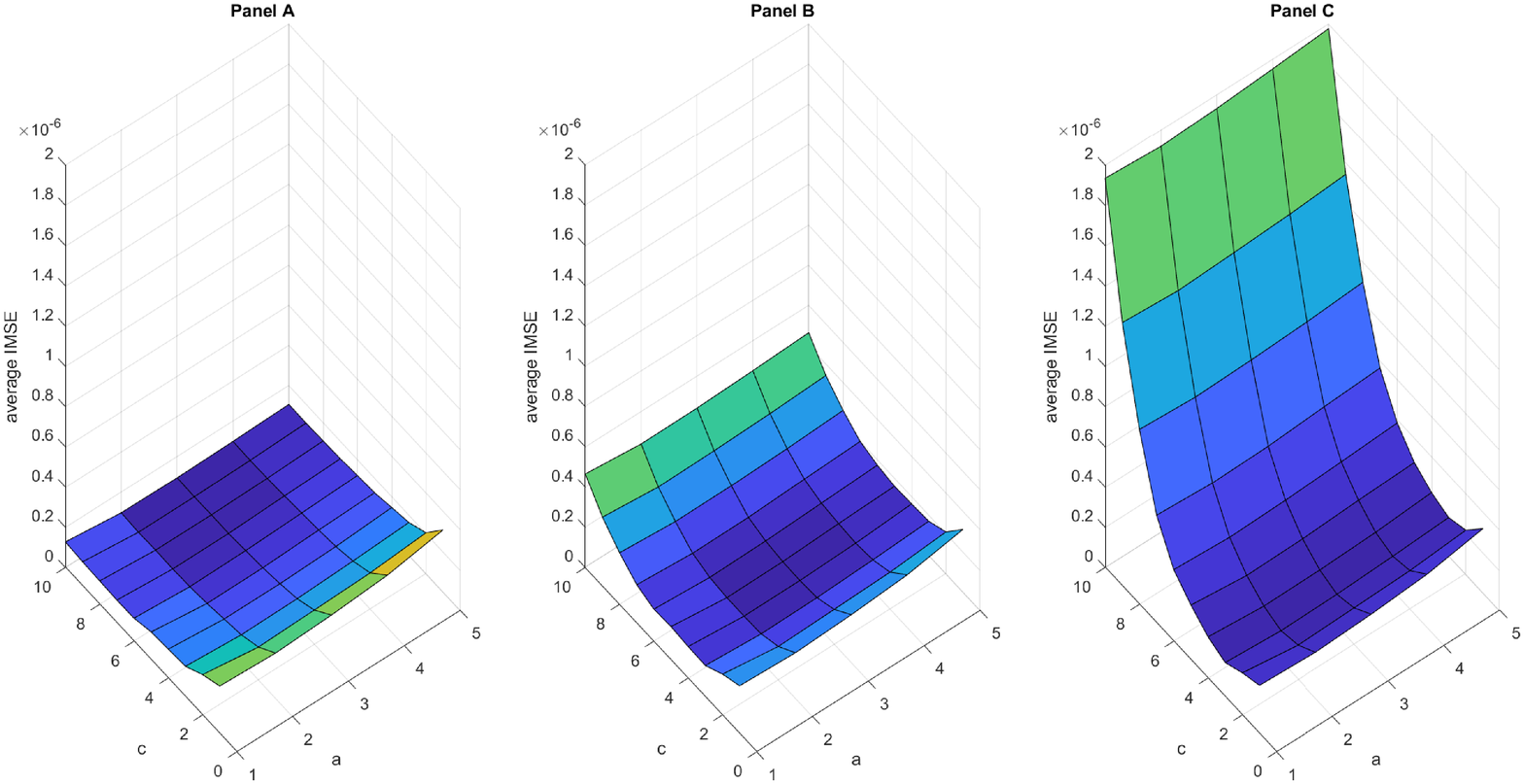}
			\caption{MIAE values over 1000 simulated trajectories of the SV1F model with $\zeta=1$ (Panel A), $\zeta=2$ (Panel B) and $\zeta=3$ (Panel C).}\label{fig:IMAEexp}
		\end{figure}
\begin{figure}[h!]
			\includegraphics[width=\textwidth]{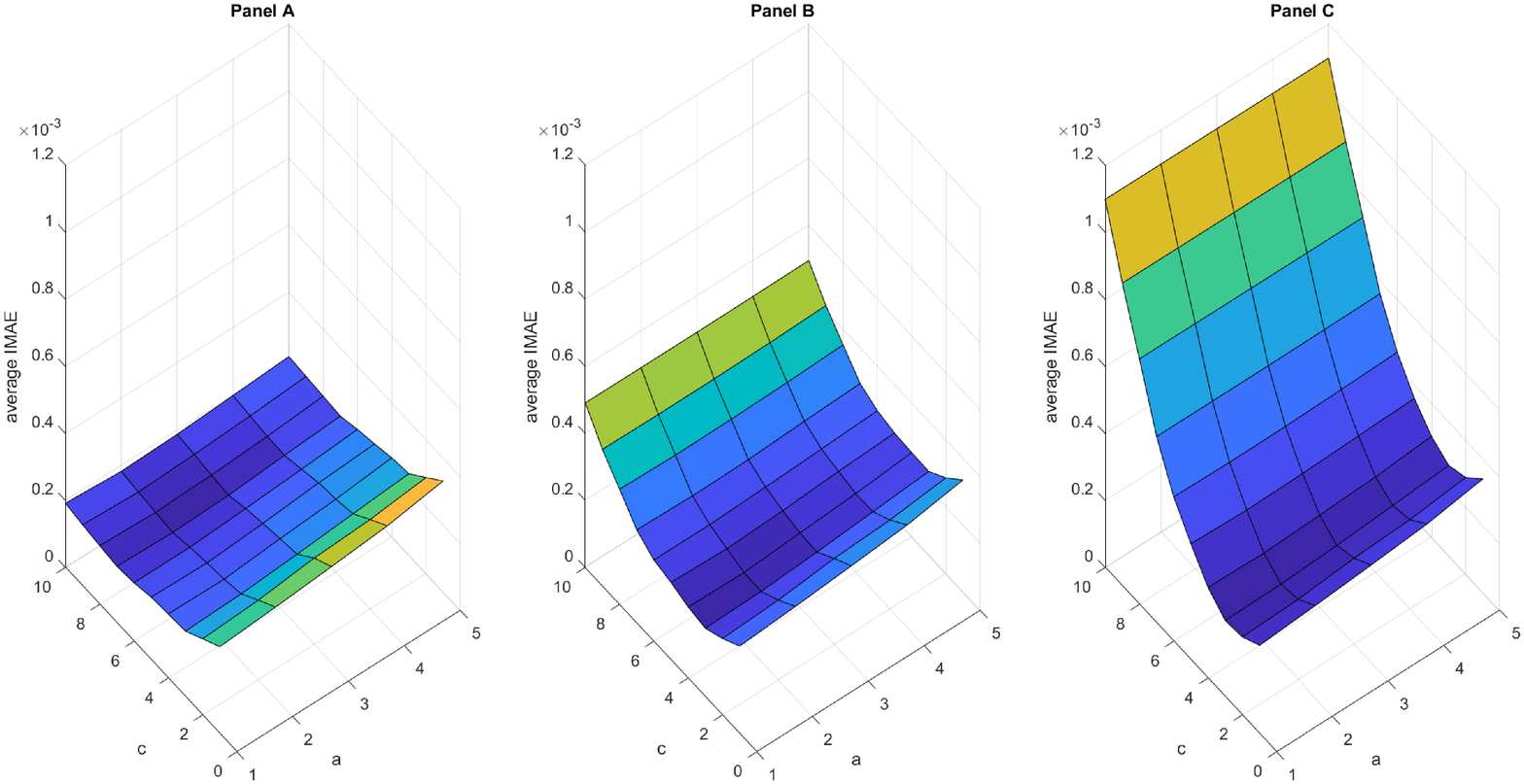}
			\caption{MIAE values over 1000 simulated trajectories of the Heston model with $\zeta=1$ (Panel A), $\zeta=2$ (Panel B) and $\zeta=3$ (Panel C). } \label{fig:IMAEhes}
		\end{figure}

		\noindent
		Figures \ref{fig:IMSEexp} - \ref{fig:IMAEhes} show the values of the MISE and MIAE as a function of the pair $({c},{a})$ and for the three different intensities of the noise-to-signal ratio considered.  Moreover, Tables \ref{table:optcaIMSE} and \ref{table:optcaIMSA}  contain the  optimal combinations of the two constants, that is, the combinations that minimize, resp., the MISE  and the MIAE. It is worth noticing that the optimal choice of $c$ gets smaller in correspondence of a larger noise-to-signal ratio.
		This is line with the fact that,  when the amount of noise present in the data is larger, a smaller value of the cutting frequency $N$ is needed in the convolution in order to efficiently filter out the noise and reconstruct the volatility coefficients. Instead, the optimal selection of $a$ appears to be  rather insensitive to changes of the noise-to-signal ratio and, moreover, MISE and MIAE values seem to be rather stable w.r.t. changes of $a$ in the region considered, for each value of the noise-to-signal ratio considered.
		
		
		\newpage
		
		\newpage
		
		\begin{table}[h!]
			\centering
			\begin{tabular}{c |  c c | c c}
				\multicolumn{1}{  c|}{ } & \multicolumn{2}{ c| }{SV1F model } & \multicolumn{2}{ c }{Heston model}
				\\
				\hline \hline
				noise-to-signal $\zeta$  & optimal $(c,a)$ &   MISE  & optimal $(c,a)$ &   MISE \\
				\hline
				$1$	&	$	(7,0.2)	$	&	$	2.490 	\cdot	10^{-7}	$	&	$	 	(9,0.2)$	&	$	 9.082	\cdot	10^{-8}	$\\
				$2$  	&	 $	(3,0.1)$ 	&	 $	2.595	\cdot	10^{-7}	$ 	&	 $	(4,0.2)	$ 	&	 $	1.376	 \cdot	10^{-7}	$ \\
				$3$  	&	 $	(2,0.1)	$ 	&	 $	2.944	\cdot	10^{-7}	$ 	&	 $	(3,0.2)	$ 	&	 $	1.513	 \cdot	10^{-7}	$ \\ 							
			\end{tabular} 
			\caption{MISE-optimal selection of the pair  $(c,a)$ for different values of the noise-to-signal ratio and  resulting MISE values.}
			\label{table:optcaIMSE}
		\end{table}
		
		\begin{table}[h!]
			\centering
			\begin{tabular}{c |  c c | c c}
				\multicolumn{1}{  c|}{ } & \multicolumn{2}{ c| }{SV1F model } & \multicolumn{2}{ c }{Heston model}
				\\
				\hline \hline
				noise-to-signal $\zeta$  & optimal $(c,a)$ & MIAE  & optimal $(c,a)$ & MIAE \\
				\hline
				$1$&	$	(7,0.2)	$	&	$	1.638 	\cdot	10^{-4}	$	&	$	(7,0.3)$	&	$	1.955	 \cdot	10^{-4}	$	\\
				$2$	&	$	(3,0.1)	$	&	$	  2.036	\cdot	10^{-4}	$	&	$	(4,0.3)	$	&	$	2.322	 \cdot	10^{-4}	$	\\
				$3$	&	$	(3,0.1)	$	&	$2.267\cdot	10^{-4} $	&	$	(3,0.3)	$	&	$	2.552	 \cdot	10^{-4}	$	\\ 							
			\end{tabular}
			\caption{MIAE-optimal selection of the pair  $(c,a)$ for different values of the noise-to-signal ratio and  resulting MIAE values.}
			\label{table:optcaIMSA}
		\end{table}

		\subsection{Feasible selection of $N$ and $M$}\label{feas}

		In this subsection  we   introduce an adaptive method for the feasible selection of the pair $(N,M)$ in finite-sample applications.  The method   aims at optimizing  the   conditional  MISE (c-MISE) of the Fourier spot volatility estimator  under the constraint  $M<N<n$, for the given available sample size $n$.
		
		Define the c-MISE as
		$$c\textit{-}MISE:= E^\sigma\left[\int_0^T\Big(\widetilde \sigma^2_{n N M}(t)- \sigma^2(t) \Big)^2  dt  \right] ,$$
		where $E^\sigma$ denotes the conditional expectation w.r.t. the filtration generated by  $(\sigma_t)_{t \in [0,T]}$. As pointed out in \cite[Section 3.4]{ZuBo},  the following decomposition holds:
		$$c\textit{-}MISE=\int_0^TE^\sigma\left[ \widetilde \sigma^2_{n N M}(t)- \sigma^2(t)    \right]^2 dt + \int_0^T Var^\sigma\left[  \widetilde\sigma^2_{n N M}(t)- \sigma^2(t)    \right] dt , $$
		where $Var^\sigma$ denotes the conditional variance w.r.t. the filtration generated by  $(\sigma_t)_{t \in [0,T]}$. Under the assumptions of the rate-efficient Central Limit Theorem in (\ref{EFm}), and after scaling the unit of time,  for a sufficiently large $n$ the   c-MISE is approximated  by  the conditional  asymptotic integrated  mean squared error (c-AMISE), which reads
		
		\begin{equation}\label{cAMISE}
			c\textit{-}AMISE = \frac{M }{N} \, {2\over 3} \,\int_0^{T} \sigma^4(t) \, dt + \frac{1}{  M} \,{T \over 3}  \, \int_0^{T} \gamma^2(t) \, dt   +   \frac{M \, N}{n} \,  {T  \over 9}\,\xi  \, \int_0^{T}\sigma^2(t) \, dt  +  \frac{N^3 M}{n^2} \, { T^3\over {15}} \,  \xi^2   .
		\end{equation}
		
		The c-AMISE in (\ref{cAMISE}) depends on the integrated volatility  $IV_T:=\int_0^T \sigma^2(t) dt$, the integrated quarticity  $IQ_T:=\int_0^T \sigma^4(t) dt $, the integrated volatility-of-volatility  $IVV_T:=\int_0^T \gamma^2(t) dt $   and the noise variance $\xi$. All these quantities can be estimated with high-frequency prices and thus the c-AMISE can be treated as observable.  Specifically, for estimating $IV_T$, $IQ_T$ and $IVV_T $ we propose to use  Fourier-transform based estimators.  We refer the reader to, resp., \cite{MS2008}, \cite{MS2012} and \cite{scm}  for the study of the finite-sample efficiency of the Fourier estimators of $IV_T$, $IQ_T$ and $IVV_T $ in the presence of noise and guidance on the practical implementation thereof with  high-frequency prices. Moreover, for estimating $\xi$, we suggest to use the estimator studied in \cite[Section 3]{BR05}.

		Let $n$ be fixed and, for brevity, denote the c-AMISE   in (\ref{cAMISE}) as a function of $(N,M)$  by $\Psi(N,M)$. Moreover,  let $\widehat \Psi(N,M)$ indicate the estimator obtained by plugging  the Fourier  estimators of  $IV_T$, $IQ_T$ and $IVV_T $, along with the estimator of $\xi$ by \cite{BR05}, into (\ref{cAMISE}), in place of the respective true latent quantities. Then, the feasible selection of $(N,M)$ for a given $n$ amounts to solving the  constrained optimization problem
		$$\displaystyle \min_{(N,M) \in S} \,  \widehat \Psi(N,M), $$
		where $S= \left[ \underline{N}, \overline{N}\right]  \times \left[ \underline{M}, \overline{M}\right]  \subset \mathbb{R}^+ \times \mathbb{R}^+ $ is such that  $M<N<n $. To do so, we propose to use the following adaptive method, based on the gradient-descent algorithm (see, e.g.,  \cite{Ruder}).

		\begin{tcolorbox}[colframe=white]
			\textbf{Adaptive method for the selection of  the pair $(N,M)$}.\\

			Given the initial condition $(N_0=\underline{N},M_0=\underline{M})$, for $k=1,...,K$,  follow the update rule\footnote{The notation $\frac{\partial \, \widehat\Psi}{\partial x}(u,v)$ is shorthand for $\frac{\partial \, \widehat\Psi(N,M)}{\partial x}\Big|_{N=u,M=v}$.}
			$$ N_{k}=  \left( \underline{N} \, \vee \, N_{k-1} -\lambda \displaystyle\frac{\partial \, \widehat \Psi  }{\partial N}  (N_{k-1},M_{k-1}) \right) \wedge \overline{N},$$
			$$ M_{k}=  \left( \underline{M} \, \vee \, M_{k-1} -\lambda \displaystyle\frac{\partial \, \widehat \Psi }{\partial M}  (N_{k-1},M_{k-1}) \right) \wedge \overline{M},$$
			where $\lambda$ is a  positive constant, typically referred to as the $learning$ $rate$ parameter.
			The optimal value $(N^*,M^*)$ is achieved  as soon as the marginal absolute change of the objective function falls below a given threshold $\vartheta$, i.e., in correspondence of the smallest $k$ such that
			$$ \delta_k:= \frac{ \left|\widehat\Psi (N_{k},M_{k})- \widehat\Psi (N_{k-1},M_{k-1})\right|}{\widehat\Psi (N_{k-1},M_{k-1})} < \vartheta. $$
			Instead, if $\delta_k>\vartheta \, \, \, \forall k$,   make the final selection   $(N^*,M^*)=(N_K,M_K).$
		\end{tcolorbox}

		The implementation of the adaptive  method  for optimizing the pair  $(N,M)$ requires the selection of the constraint region $S$ and the vector of optimization parameters $(\lambda, \vartheta, K)$. In this regard, given the sample size $n=23400$, corresponding to the sampling frequency $\rho(n)=1$ second, we set
		\begin{equation}
			\label{finetuning1}
			S = \left[ \lfloor  {2}^{-1} n^{1/2} \rfloor, \lfloor 10 \, n^{1/2} \rfloor \right] \times \left[\lfloor {10}^{-1} \, n^{1/4} \rfloor, \lfloor 2 \, n^{1/4} \rfloor \right]
		\end{equation}
		and
		\begin{equation}\label{finetuning2}
			(\lambda, \vartheta, K)= (500 \, \hat \xi ^{-1}, 10^{-3}, 10^5).
		\end{equation}

		Note that the choice of $S$ was performed rather conservatively, by allowing a fairly large range of variation for $N$ and $M$. Analogously, $\vartheta$ and $K$ were chosen conservatively to ensure that, trajectory-wise, the stopping rule
		$\delta_k < \vartheta$ was satisfied for some $k\le K$. Moreover, particular attention was devoted to the choice of the learning rate $\lambda$, as   the latter   plays a crucial role in determining the accuracy of the gradient-descent algorithm, see \cite{Ruder} and the references therein. In this regard,   simulations suggest that a value  inversely proportional to the (estimated) variance of the noise yields satisfactory results. In particular, after setting $$\lambda=c_\lambda \, \hat \xi ^{-1},$$  we numerically evaluated the efficiency of the adaptive method for different values of $c_\lambda$ in the region $[1,1000]$ and found that  the MISE and MIAE values were optimized for  $c_\lambda$ in a neighborhood of $500$,
		both in the case of the SV1F model and the Heston model, hence the selection of $\lambda$ in (\ref{finetuning2}). As for robustness, MISE and MIAE values appeared to be rather stable for $c_\lambda \in [300, 700]$, based on numerical evidence.

		The performance of the feasible adaptive method in terms of MISE and MIAE is summarized in Table \ref{table:adapt}. Note that the  latter  is quite satisfactory, compared to the results of the unfeasible optimization exercise conducted in Subsection \ref{sect:sensitivity} (see Tables \ref{table:optcaIMSE} and \ref{table:optcaIMSA}). In fact, MISE and MIAE values are smaller in Table \ref{table:adapt}. The better performance of the feasible adaptive method is explained by the fact that it optimizes the pair $(N,M)$ trajectory-wise.

		\begin{table}[h!]
			\centering
			\begin{tabular}{c |  c c | c c}
				\multicolumn{1}{  c|}{ } & \multicolumn{2}{ c| }{SV1F model } & \multicolumn{2}{ c }{Heston model}
				\\
				\hline \hline
				noise-to-signal $\zeta$   & MISE & MIAE  &   MISE   &   MIAE \\
				\hline
				$1$ &	 $	9.547 \cdot	10^{-8}	 $  	&	$	1.216 \cdot	10^{-4}		$	&	$	8.882 \cdot 10^{-8} 	 $	&	$	 	1.913\cdot	10^{-4}	$	\\
				$2$  	&	 $	1.053 \cdot 10^{-7} $ 	&	 $ 1.335 \cdot 10^{-4}$ 	&	 $	 1.312 \cdot 10^{-7}	$ 	&	  $	2.300	\cdot	10^{-4}	 	$ 	\\
				$3$  	&	 $	1.439 \cdot 10^{-7} 	$ 	&	 $1.480 \cdot 10^{-4}$ 	&	 $ 1.447		\cdot	10^{-7}	 $ 	&	 $2.507 \cdot	10^{-4}	$ 	\\ 							
			\end{tabular} 
			\caption{MISE and MIAE values produced by the feasible adaptive method for different values of the noise-to-signal ratio.}
			\label{table:adapt}
		\end{table}
		\begin{Remark}
			As pointed out in \cite[Section 3.4]{ZuBo}, the c-MISE is a  global measure of  path-wise efficiency over the entire estimation interval. Accordingly, the adaptive method introduced in this section to minimize the c-MISE yields as an output a  path-wise  optimal value of the pair $(N,M)$ which is independent of the  time instant $t	\in [0,T]$. The results of the simulation study conducted in this section thus offer support to the efficiency of the Fourier estimator as a global estimator of the spot volatility path. This is line with the intrinsic global character of the Fourier spot volatility estimator, see \cite{MM09}. As already remarked in \cite{Plos}, the global efficiency of the Fourier estimator may be relevant for financial applications (e.g., for the calibration of stochastic volatility models).
		\end{Remark}
		
		\subsection{Performance comparison with alternative noise-robust estimators}
		
		This subsection illustrates a comparison of the finite-sample performance of the Fourier spot volatility estimator  with that of alternative   noise-robust spot volatility estimators, namely the Two-scale estimator by \cite{ZuBo} and the Pre-averaging kernel estimator by \cite{FigueroaWu}. We briefly recall the definitions of the two estimators.
		
		The Two-scale   estimator  by \cite{ZuBo} is defined as
		\begin{equation}
			\label{TS}
			\widetilde{\sigma}_{k h}^2(t):=\frac{1}{h}\sum_{t\le t_i\le t+h}\frac{(\widetilde{p}_{t_{i+k}}-\widetilde{p}_{t_{i}})^2}{k}  -\frac{\bar n}{n}\frac{1}{h}\sum_{t\le t_i\le t+h}  \delta_i (\widetilde{p})^2,
		\end{equation}
		where $\bar n = \frac{nh-k+1}{kh}$, $k=\lfloor c_k n^{2/3}\rfloor$ and $h= c_h n^{-1/6}$. The implementation of (\ref{TS}) then requires the selection of the constants $c_k$ and $c_h$. For this task, we followed \cite[Section 3.4]{ZuBo}.

		The pre-averaging kernel estimator by \cite{FigueroaWu} reads
		
		\begin{equation}\label{PAV}
			\widetilde{\sigma}_{k m}^2(t):=\frac{1}{\phi_k(g)}\sum_{i=1}^{n-k+1}\frac{1}{m\Delta_n}K\left(\frac{t_{i}-t}{m\Delta_n}\right)\left(\bar{P}_i^2-\frac{1}{2}\widehat{P}_i\right),
		\end{equation}
		where: $\Delta_n=T/n$, $k=\lfloor 1/\left(c_k \sqrt{\Delta_n} \right) \rfloor$ and $m=c_m \Delta_n^{-3/4}$; $K$ is a kernel function satisfying  Assumption 2 of \cite{FigueroaWu}; $g$  is a real-valued weight function on $[0,1]$ continuous and piece-wise $C^1$ with a piece-wise Lipschitz derivative $g^{\prime}$; moreover $g(0)=g(1)=0$, $\int_0^1 g^2(s)ds =1$; finally,
		$$\bar{P}_i = -\sum_{j=1}^{k}\left(g\left(\frac{j}{k}\right)-g\left(\frac{j-1}{k}\right)\right)\widetilde{p}_{i+j-2}, \quad \widehat{P}_i=\sum_{j=1}^{k}\left(g\left(\frac{j}{k}\right)-g\left(\frac{j-1}{k}\right)\right)^2\delta_{i+j-2} (\widetilde{p})^2, \quad\phi_k(g) =\sum_{j=1}^{k}g\left(\frac{j}{k}\right)^2.$$
		To apply the estimator in (\ref{PAV}), one then needs to select a kernel function $K$ and a pre-averaging function $g$, along with the constants $c_k$ and $c_m$.  Based on the discussion in \cite[Section 3]{FigueroaWu},  we selected $K(x)=\frac{1}{2}\textrm{e}^{-|x|}$. Further, we chose $c_k$ and $c_m$ in accordance with the formulas derived by the authors to optimize the integrated asymptotic variance (see also Remark 4.1 in \cite{FigueroaWu}). As for the function $g$, we selected $g(x) =  x \wedge (1-x)$, as in \cite{pavorig}. 
		
		Tables \ref{table:TS} and \ref{table:PAVK} summarize the results of the comparative performance study.
		\begin{table}[h!]
			\centering
			\begin{tabular}{c |  c c | c c}
				\multicolumn{1}{  c|}{ } & \multicolumn{2}{ c| }{SV1F model } & \multicolumn{2}{ c }{Heston model}
				\\
				\hline \hline
				noise-to-signal $\zeta$   &  MISE &  MIAE  &  MISE   &   MIAE \\
				\hline
				$1$ &	 $4.986	  \cdot	10^{-7}	 $  	&	$	  3.297  \cdot	10^{-4}		$	&	$	3.951  \cdot 10^{-7} 	 $	&	$	 5.247 	 \cdot	10^{-4}	$	\\
				$2$  	&	 $	  7.016  \cdot 10^{-7} $ 	&	 $   3.304 \cdot 10^{-4}$ 	&	 $	9.326   \cdot 10^{-7}	$ 	&	 $   7.071	\cdot	10^{-4}	 	$ 	\\
				$3$  	&	 $	8.045  \cdot 10^{-7} 	$ 	&	 $ 2.864 \cdot 10^{-4}$ 	&	 $  1.179	\cdot	10^{-6}	 $ 	&	 $ 7.777  \cdot	10^{-4}	$ 	\\ 							 
			\end{tabular} 
			\caption{Two-scale estimator (\ref{TS}): MISE and MIAE values  for different values of the noise-to-signal ratio.}
			\label{table:TS}
		\end{table}
		
		\begin{table}[h!]
			\centering
			\begin{tabular}{c |  c c | c c}
				\multicolumn{1}{  c|}{ } & \multicolumn{2}{ c| }{SV1F model } & \multicolumn{2}{ c }{Heston model}
				\\
				\hline \hline
				noise-to-signal $\zeta$ &   MISE &  MIAE  &  MISE   &  MIAE \\
				\hline
				$1$ &	 $	2.146  \cdot	10^{-7}	 $  	&	$	1.898  \cdot	10^{-4}		$	&	$	1.928  \cdot 10^{-7} 	 $	&	$	3.105 	 \cdot	10^{-4}	$	 \\
				$2$  	&	 $	2.379  \cdot 10^{-7} $ 	&	 $  1.916 \cdot 10^{-4}$ 	&	 $	1.944   \cdot 10^{-7}	$ 	&	 $ 3.176	\cdot	10^{-4}	 	$ 	\\
				$3$  	&	 $ 2.382	  \cdot 10^{-7} 	$ 	&	 $  1.939 \cdot 10^{-4}$ 	&	 $ 2.191	\cdot	10^{-7}	 $ 	&	 $ 3.366 \cdot	10^{-4}	$ 	\\ 							 
			\end{tabular} 
			\caption{Pre-averaging kernel estimator (\ref{PAV}): MISE and MIAE values for different values of the noise-to-signal ratio. }
			\label{table:PAVK}
		\end{table}
		
		The comparison of Tables \ref{table:TS} and \ref{table:PAVK}  with Table \ref{table:adapt} suggests that the Fourier estimator yields  the lowest MISE and MIAE values under both parametric settings adopted and for each of the three noise-to-signal ratios employed. Moreover, the Pre-averaging estimator appears to produce lower MISE and MIAE values compared to the Two-scale estimator in each scenario considered. This is coherent with the numerical results illustrated in \cite[Section 4.6]{FigueroaWu}. However, it is  worth noticing that the performance gap observed in the simulation study between the Fourier and the Pre-averaging estimators  appears to decrease as the noise-to-signal ratio is increased, in terms of both MISE and MIAE.

		Overall,  we observe that the numerical findings of this subsection robustify the numerical evidence produced in \cite{Plos} by taking into account also the estimator  by \cite{FigueroaWu} for the performance comparison. Furthermore, it is worth noticing that the performance ranking of spot volatility estimators obtained here is coherent with the ranking obtained in the simulation study by \cite{MLT}, where the data-generating process is the limit order book simulator by \cite{HLR}.
		
		\section{Empirical study}\label{emp}
		
		In this section we illustrate a study on the accuracy of the Fourier spot volatility estimator with S\&P500 high-frequency trade prices, sampled over the period March, 2018 - April, 2018.
		
		The accuracy of spot volatility estimates can be investigated empirically by testing the distribution of the reconstructed standardized log-return, see \cite[Section 4]{Plos}. Define the standardized log-return on $[t,t+h]$, $h>0$, as
		\begin{equation}\label{z}
			z_h(t):=\frac{r_h(t)}{\sqrt{\sigma^2(t)\, h}},
		\end{equation}
		where $r_h(t):=p(t+h)-p(t)$. Moreover, denote  by $z_h^{F}(t)$ the reconstructed standardized log-return, obtained by replacing $\sigma^2(t)$ in (\ref{z}) with the Fourier estimate thereof. Under Assumption (A.I), if the interval $h$ is sufficiently small, the sequence $\{z_h(t)\}_{t \in \tau}$, $\tau=\{0,h, 2h, ...,  (\lfloor T/h \rfloor-1)h \}$, approximates a sequence of i.i.d. standard Normal random variables. By testing the empirical distribution of the reconstructed standardized log-returns $\{z^F_h(t)\}_{t \in \tau}$ for Normality, one can then obtain insight on the accuracy of Fourier spot volatility estimates.
		
		The estimation of the spot volatility was conducted on the daily horizon, that is, on the time interval between 9:30  a.m. and 4:00  p.m. As we measured time in days, the  corresponding value of $T$ was set equal to $1$.  For the implementation of the Fourier   estimator, we used the sampling frequency $\rho(n)=1$ second, corresponding to $n=23400$. As for the selection of $N$ and $M$, we employed the adaptive method described in  Subsection \ref{feas} and, based on the recommendation from the simulation study, we made the choice of the rectangle $S$ and the optimization parameters $\lambda, \vartheta, K$ as in, respectively,  (\ref{finetuning1}) and (\ref{finetuning2}). Note that, on average, the adaptive method lead to the selection of $N=305.48$ and $M=6.94$.

		Days with price jumps were not considered. In this regard, after applying the jump test by \cite{LM08} at the 95\% confidence level, we detected the presence of intraday jumps only on March 14th and thus the corresponding daily sub-sample  was discarded. Overnight returns, that is, the difference between the opening log-price and the closing log-price of the previous day, were not employed in the estimation, as they typically contain jumps due to the fact that news releases usually take place when the market is closed.  In fact, the test by \cite{LM08}  at the 95\% confidence level  detected jumps in correspondence of overnight returns for approximately 40\% of the days of the 2-month sample object of study. Overall, we obtained $41$ spot volatility daily paths.
		
\begin{figure}[h!]
			\includegraphics[width=\textwidth]{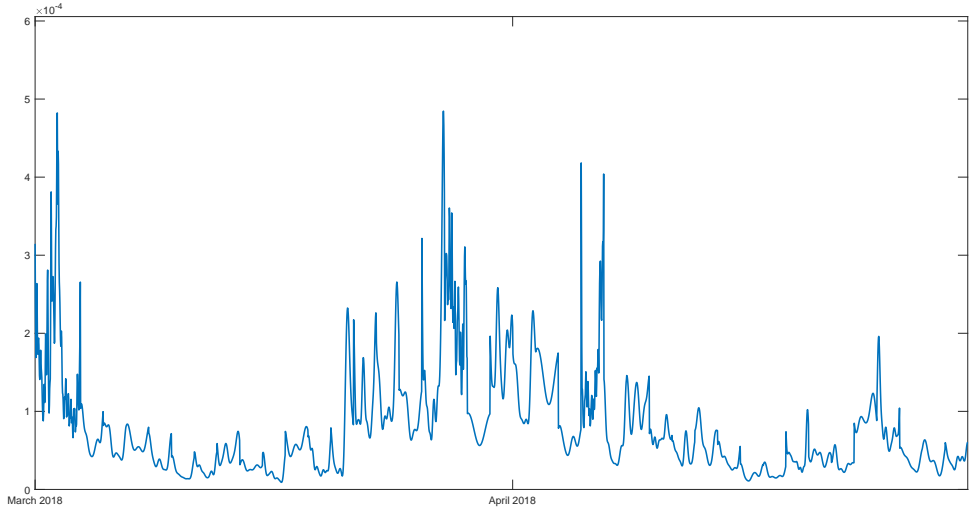}
			\caption{S\&P500 (March, 2018 - April, 2018): sequence of the reconstructed daily volatility paths on the 5-minute grid. } \label{fig:volpath}
		\end{figure}

		In order to test the distribution of the reconstructed standardized log returns, one needs to select a value for the interval $h$ which is, at the same time, small enough to satisfactorily approximate the standard Gaussian distribution  and large enough to make the impact of micro-structure noise negligible. Based on the application of the noise-detection  test by \cite{AX}, we selected $h$ equal to $5$ minutes. 
		Accordingly, Fourier estimates of the spot volatility were obtained on the equally-spaced grid with mesh size equal to 5 minutes. However, it is worth stressing that the Fourier method allows selecting an arbitrary grid  for the estimation of the volatility path and that the selection of the 5-minute grid is only convenient for the purpose of the testing of the reconstructed standardized log-return distribution.  The  sequence of the reconstructed daily volatility paths is displayed in Figure \ref{fig:volpath}.
		
		We applied the Jarque-Bera test (see \cite{JB}) to the $41$ daily samples of reconstructed standardized log-returns. At the 95\% confidence level, the null hypothesis of Gaussianity was never rejected.  The average p-value was equal to $0.374$. Moreover, Table \ref{table:emp} reports the averages and standard deviations, over the 41 daily samples analyzed, of the sample mean, variance, skewness and kurtosis of the reconstructed standardized log-returns. The averages are very close to the theoretical moments of a standard Normal, while the standard deviations are relatively small. Overall, the results of the Jarque-Bera test, along with the empirical evidence on the sample  moments, suggest that the daily empirical distributions of the reconstructed standardized log-returns approximate a standard normal quite well, thereby supporting the accuracy of the Fourier spot volatility estimates obtained.
		\begin{table}[h!]
			\centering
			\begin{tabular}{c | c c   } 
				&   average  &   std. dev.     \\
				\hline \hline
				mean 	&  -0.005 & 0.132   \\
				variance 	&  1.008   & 0.182   \\
				skewness 	& 0.006  &  0.222  \\
				kurtosis 	& 2.940  &  0.268  \\		\end{tabular} 
			\caption{S\&P500 (March, 2018 - April, 2018):  averages and standard deviations, over the 41 daily samples analyzed, of the sample statistics of the reconstructed standardized log-returns.}
			\label{table:emp}
		\end{table}

		\section{Conclusions}
		\label{concl}
		
		We have proved two rate-optimal central limit theorems for the Fourier estimator of the spot volatility both in the absence and in the presence of additive microstructure noise. In the absence of noise, our result completes the asymptotic theory presented in \cite{Plos}, where the rate is slightly sub-optimal under more general hypotheses on the volatility process.
		Additionally, we have derived a feasible adaptive method for the optimal selection of the parameters involved in the implementation and provided numerical results which support its accuracy. 
		Finally, we offered support to the reliability of the adaptive method for practical applications via an empirical study, conducted with a short sample of  S\&P500 high-frequency prices.
		
		\newpage
		\section{Appendix A}
		\label{Proof}
			
		\textit{Please contact the authors for the proofs.}
		
		\section{Appendix B}
		\label{kernelproperties}
		
		This Appendix resumes some results about the rescaled Dirichlet kernel, defined as
		\begin{equation}
			\label{dirichlet}
			D_N(x):= {1\over {2N+1}} \sum_{|k|\leq N} e^{{\rm i}kx} = {1\over {2N+1}} {\sin ((2N+1)x/2) \over {\sin (x/2)}}
		\end{equation}
		and the Fej\'{e}r kernel, defined as
		\begin{equation}
			\label{FejerDef}
			F_M(x):= \sum_{|k| \leq M}\left(1-{|k|\over {M+1}}\right) \, e^{{\rm i}kx}={1 \over{M+1}}\left({\sin((M+1) x/2) \over {\sin(x/2)}}\right)^2.
		\end{equation}

		\begin{Lemma}
			\label{LemmaFejer}
			i) For any $M$,
			$$\int_{-\pi}^{\pi} F_M(x)dx=2\pi.$$
			Moreover, under the assumption $\lim_{n,M\to \infty} {M^{\tau}\over n}=a$, for a constant $a>0$ and for some $\tau>1$, it holds:
			$$\lim_{n,M \to \infty} \int_{-\pi}^{\pi} F_M(\varphi_n(x)) dx= 2\pi$$
			and, in particular,
			$$\left|\int_{-\pi}^{\pi} F_M(\varphi_n(x)) dx- \int_{-\pi}^{\pi} F_M(x) dx \right| \leq C {M \over n}.$$
			ii) For any $M\geq 1$,
			$$ \int_{2\pi \over {M+1}}^{\pi} F_M(x)dx \leq {C \over M}.$$
			iii) Under the assumption $\lim_{n,M\to \infty} {M^{\tau}\over n}= a$, for a constant $a>0$ and for some $\tau>1$,
			it holds:
			\begin{equation}
				\label{squarefejerTEIC}
				\lim_{M\to \infty} \int_{-\pi}^{\pi} {1\over M} F^2_M(x) dx =
				\lim_{M,n\to \infty} \int_{-\pi}^{\pi} {1\over M} F^2_M(\varphi_n(x)) dx = {4\pi \over 3}.
			\end{equation}
			iv)
			If $\lim_{n,M\to \infty} {M^{\tau}\over n}= a$, for a constant $a>0$ and  for some $\tau>1$, and $f$ is H\"older continuous with parameter $\alpha \in (0,1]$, then
			\begin{equation}
				\label{Key2}
				\lim_{n,M\to \infty}{{1\over M}} \int_{-\pi}^{\pi}  F_M^2(x-\varphi_n(y)) f(y) dy = \lim_{M\to \infty} \int_{-\pi}^{\pi} {1\over M} F_M^2(x-y) f(y) dy=
				{4\pi\over 3}f(x).
			\end{equation}
		\end{Lemma}
		\Proof See \cite[Lemma 5.1]{CuTe}.

		\begin{Remark}
			\label{trasla}
			As observed in \cite{CuTe} Remark 5.2, all above results similarly hold true on $[0,T]$.
		\end{Remark}

		\begin{Lemma}
			\label{derivaF}
			Suppose that $ \lim_{n,M \to \infty}{M^{\tau}\over n}= a$,  for some constant $a > 0$ and $\tau>1$. Then, it holds that:
			\begin{equation}
				\label{eq:first_derivative_square}
				\lim_{n, M\to \infty} \int_{-\pi}^{\pi}\frac{1}{M^3}|F^{\prime}_M(\varphi_n(x))|^2\,dx = \lim_{M\to \infty} \int_{-\pi}^{\pi}\frac{1}{M^3}|F^{\prime}_M(x)|^2\,dx = \frac{2}{15}  \pi,
			\end{equation}
			\begin{equation}
				\label{eq:second_derivative_square}
				\lim_{n, M\to \infty } \int_{-\pi}^{\pi}\frac{1}{M^5} |F^{\prime\prime}_M(\varphi_n(x))|^2\,dx  = \lim_{M \to \infty} \int_{-\pi}^{\pi} \frac{1}{M^5} |F^{\prime\prime}_M(x)|^2\,dx = \frac{4}{105} \pi.
			\end{equation}
		\end{Lemma}
		\Proof See \cite[Lemma 8.1]{TLMM}.

		\begin{Lemma}
			\label{Key}
			i) If $\lim_{n,N \to \infty} {N\over n}=c > 0$, then for any $p>1$ there exists a constant $C_p$ such that
			$$\lim_{n,N \to \infty}\, n \sup_{x\in [0,2\pi]} \int_0^{2\pi} D^p_N(\varphi_n(x)-\varphi_n(y)) dy \leq C_p.$$
			ii) If $\lim_{n,N \to \infty} {N\over n}=c > 0$, it holds that
			$$
			\lim_{n,N\to\infty }n \, \int_0^{x} D^2_{N}(\varphi_n(x)-\varphi_n(y)) dy = \pi(1+2K(2c))
			$$
			and, for any $\alpha$-H\"older continuous function $f$ with $\alpha \in (0,1]$,
			$$
			\lim_{N,n\to \infty}n \int_0^{x} D^2_{N}(\varphi_n(x)-\varphi_n(y)) f(y) dy = \pi (1+2K (2c)) \, f(x),
			$$
			where
			$$
			K(c):= \frac{1}{2c^2}r(c)(1-r(c)),
			$$
			being $r(c)=c-[c]$, with $[c]$ denoting the integer part of $c$.
			\par\noindent
			iii) If $\lim_{n,N \to \infty} {N\over n}=c > 0$, then for any $s\in [0,2\pi)$ and for any $\varepsilon >0$,
			$$\lim_{n,N \to \infty} \ n\int_0^{x-\varepsilon} D^2_N(\varphi_n(x)-\varphi_n(y)) dy =0.$$
		\end{Lemma}
		\Proof See \cite[Lemma 3]{ClGl}.
		
		\begin{Lemma}
			\label{Key1}
			If $\lim_{N,n\to \infty}{N^{\tau} \over n}= c$, for a constant $c>0$ and some $\tau>1$, then
			$$
			\lim_{n,N \to \infty}N \, \int_0^{x} D^2_{ N}(\varphi_n(x)-\varphi_n(y)) dy  = \lim_{N \to \infty}{ N} \, \int_0^{x} D^2_{ N}(x-y) dy
			={\pi \over 2}.
			$$
		\end{Lemma}
		\Proof See \cite[Lemma 5.1]{CuTe}, using the fact that $D_N^2(x)= (2N+1)^{-1}F_{2N}(x)$.

		\begin{Lemma}
			\label{derivaD}
			\par\noindent
			Under the condition $\lim_{N,n\to \infty}{N^{\tau} \over n}= c$, for a constant $c>0$ and some $\tau>1$, it holds:
			$$i) \,\,\, \lim_{n,N\to \infty }{1\over N}\int_0^{2\pi}| D^{\prime}_{N}(\varphi_n(x))|^2 dx=\lim_{N\to \infty} {1\over N}\int_0^{2\pi} |D^{\prime}_N(x)|^2 dx=  {\pi\over 3}.
			$$
			$$ii) \, \, \, \lim_{n,N\to \infty}{1\over {N^3}}\int_0^{2\pi} |D^{\prime\prime}_N(\varphi_n(x))|^2 dx =\lim_{N\to \infty}{1\over {N^3}}\int_0^{2\pi} |D^{\prime\prime}_N(x)|^2 dx= {\pi \over 5}.$$
		\end{Lemma}
		\Proof
		For i) see \cite{MaToLev}. As for ii), it holds that:
		$$\frac{1}{N^3}\int_0^{2\pi} \left| D^{\prime\prime}_N(x)\right|^2 dx =\frac{1}{N^3} \frac{2\pi}{\left(2N+1\right)^2} \sum_{|k|\le N}k^4 =\frac{2\pi}{15}  \frac{N(N+1)(3N^2+3N-1)}{N^3(2N+1)}  \rightarrow \frac{\pi}{5}.$$

	\end{document}